\documentclass[twocolumn]{aastex631}
\usepackage{bm}
\usepackage{amsmath}

\usepackage{orcidlink}
\graphicspath{{./}{figures/}}

\usepackage{hyperref}
\begin{document}

\title{Stability of circumbinary orbits in misaligned triple star systems }

\author[0000-0003-2270-1310]{Stephen Lepp}
\author[0000-0003-2401-7168]{Rebecca G. Martin}
\affiliation{Nevada Center for Astrophysics, University of Nevada, Las Vegas, 4505 S. Maryland Pkwy., Las Vegas, NV 89154, USA}
\affiliation{Department of Physics and Astronomy,University of Nevada, Las Vegas, 4505 S. Maryland Pkwy., Las Vegas, NV 89154, USA}

\begin{abstract}

We investigate the stability of circumbinary orbits in hierarchical  triple star systems, focusing on the effects of a misaligned outer companion star. Test particles are subject to competing gravitational torques from the inner binary and the outer binary companion. With secular theory we estimate the outer radius of particle stability, where the torques balance. We find good agreement with $n$-body simulations across a wide range of triple star configurations. Stable circumbinary orbits can exist even in strongly misaligned triples. Polar and highly inclined orbits with respect to the inner binary can remain stable over a substantial radial range, which is insensitive to the triple star misalignment.  Orbits that are close to coplanar or retrograde coplanar to the inner binary are more susceptible to instability when the mutual inclination between the binaries is large.  These findings indicate that misaligned and polar circumbinary disks and planets can survive in triple star systems under a broad set of conditions. The analytic criterion identifies where stable material may exist, with implications for the formation and detection of circumbinary planets in multiple-star systems.

\end{abstract}


\section{Introduction}
\label{sec:intro}

Triple star systems are generally hierarchical since this is a  more stable configuration \citep{Mardling2001,Valtonen2006,Toonen2016,Vynatheya2022,Tokovinin2021}. Hierarchical triple stars are composed of an {\it inner binary} that is in an orbit with another companion, the  {\it outer binary}.  In a triple star system, misalignments between the inclination of the inner and outer binary ($i_{AB}>0$) may also be common, particularly for wider outer binaries \citep{Tokovinin2017}.  Large mutual inclinations including counter rotating inner and outer binaries have been observed \citep{Tokovinin2016,Schaefer2016, Mitnyan2024}. For sufficiently highly inclined orbits, the triple star may undergo Kozai-Lidov (KL) oscillations \citep{Zeipel1910,Kozai1962,Lidov1962}, where the inclination and eccentricity of the inner binary are exchanged \citep[e.g.,][]{Naoz2016}. There is a peak in the observed mutual inclination distribution at around $i_{AB}=40^\circ$ \citep{Borkovits2016}, evidence that KL oscillations are operating in triple star systems \citep{Kiseleva1998,Fabrycky2007}.

Observations show that misalignments between the orbital plane of a circumbinary disk and the inner binary orbital plane are common \citep{Czekala2019}. This may be a result of turbulence in the molecular gas cloud \citep{Bateetal2003,Offner2010, Tokuda2014, Bate2012}, later  accretion  \citep{Bate2010, Bate2018}, stellar flybys \citep{clarke1993,Cuello2019b,Nealon2020} or bound stellar companions \citep[e.g.,][]{Aly2015,Martin2017,Martin2022}.
Circumbinary disk misalignments of up to $90^\circ$ are observed. Polar gas disks have been observed  in the young systems HD 98800 \citep{Kennedy2019} and V773 Tau B \citep{Kenworthy2022}. In the polar alignment, the disk angular momentum vector is aligned to the binary eccentricity vector.  The debris disk around 99 Her shows a polar ring \citep{Kennedy2012,Smallwood2020}.    The
peculiar light curve of KH 15D can be explained by a misaligned
precessing circumbinary disk \citep[e.g.,][]{Winn2004, Chiang2004,Capelo2012, Smallwood2019,Poon2021}. The
binary protostar IRS~43 has a misalignment angle greater than $60^\circ$
between the binary and the disk \citep{Brinch2016}.

Misalignments are also observed between circumbinary disks and the outer binary companion. Both HD98800 and V773 Tau B have misalignments between the circumbinary disk and the outer binary orbit.  V892 Tau is a triple system with a circumbinary disk that is close to alignment to the inner binary orbital plane but the companion star is misaligned by about $i_{AB}=60^\circ$. \citep{Alaguero2024}.

Most observed circumbinary planets are close to coplanar to the inner binary orbit \citep{Doyle2011,Welsh2012}. This may be a result of selection effects since coplanar planets are much easier to observe
 \citep[e.g.][]{MartinDV2015,MartinDV2017,MartinDV2019}. However, a recent observation of retrograde apsidal precession of a binary \citep{Baycroft2025} suggests the presence of a polar circumbinary planet \citep[e.g.][]{Zhang2019, Childs2023,Lepp2023,Lubow2024}. Low mass polar planets may form in a polar circumbinary disk as efficiently as coplanar planets in a coplanar disk \citep[e.g.][]{Childs2021,Childs2021b}. Since circumbinary disks can have large misalignments, we expect planets may form with their orbits misaligned to an inner binary orbit and/or an outer binary orbit.

  Stability of test particle orbits have been studied in detail previously around one component of a binary \citep{Holman1999,Quarles2020}, around an isolated binary \citep{Doolin2011,Chen2020,Wang2023,deElia2026}, massive planets around an isolated binary \citep{Chen2020,Georgakarakos2024}, and more generally in eccentric and mutually inclined planetary systems \citep{Hadden2018,Bhaskar2024}. Most previous studies of test particles in triples have considered coplanar triples \citep[e.g.,][]{Verrier2007,Verrier2008,Busetti2018,Martin2022} or coplanar planets \citep{Hamers2016}.  
 We examine the stability of test particle orbits around a binary star with a misaligned outer companion. The results also have implications for circumbinary disk stability since the rings of the disk feel the same torque as the planet.  In Section~\ref{sec:dynamics} we use secular theory to explore the particle dynamics. In Section~\ref{sec:sim} we discuss the $n$-body simulation set-up.  In Section~\ref{sec:results} we present stability maps for the $n$-body simulations. In Section~\ref{sec:discuss} we discuss the relation to other work, specifically the similarity of this setup with that of a test particle orbiting around a tilted oblate planet, and the relevance to the evolution of circumbinary gas disks.
 We draw our conclusions in Section~\ref{sec:conc}

\section{Circumbinary dynamics with an outer binary}
\label{sec:dynamics}

A test particle orbiting around a binary with an outer binary companion feels competing torques from the inner and outer binary. In this Section we estimate each of these torques and find the particle orbital radius where they are equal in magnitude.

\subsection{Our standard parameters}
\label{sec:standard}

The system is scale free and so we will define all masses with respect to the mass of the inner binary, $m_A$,  and all distances with respect to the semimajor axis of the inner binary, $a_A$. The masses of the inner binary components are  $m_{Aa}$ and $m_{Ab}$ with $m_{Aa}+m_{Ab}=m_A$. For our standard parameters we take an equal mass binary with $m_{Aa}=m_{Ab}=0.5 \,m_A$. The stars orbit with an eccentricity of $e_A=0.5$.  The system has a distant companion of mass $m_B=m_A$ orbiting at a semimajor axis of $a_{AB}=30\, a_A$ in a circular orbit, $e_{AB}=0.0$.  The system can be characterized  by the ratio of the masses in the inner binary $f_A=m_{Ab}/(m_{Aa}+m_{Ab})=0.5$.  We take $m_{Aa}$ as the larger of the two masses and so $f_A$ can range from 0 to 0.5.  
 
The scale free nature of the system with respect to the size and mass is broken if additional effects become important such as tidal interactions, stellar collisions, or general relativity \citep[e.g.][]{Zanardi2018,Lepp2022,Lepp2024}. Depending on the physical scale of a particular stellar triple, some of the highly eccentric configurations considered here could instead undergo significant tidal evolution or even a stellar collision but we include them for a full description.

\subsection{Torque from the inner binary}

Depending on the initial particle inclination relative to the inner binary, the orbits can undergo circulating or librating nodal precession.   At low inclinations, the angular momentum vector of the test particle precesses about the angular momentum of the inner binary which is a circulating orbit.  At higher inclinations, the angular momentum vector of the test particle precesses about the eccentricity vector of the inner binary which is a librating orbit.  At even higher inclinations, the angular momentum vector precesses about the about the negative of the binaries angular momentum vector which is a retrograde circulating orbit.

The timescale for the nodal precession of a test particle orbiting about the inner binary with radius $r$ is given 
approximately by
\begin{equation}
    t_{\rm prec}=\frac{P_A m_A^2}{k\, m_{Aa} m_{Ab}} \left(\frac{r}{a_A}\right)^{7/2},
\end{equation}
where $P_A =2\pi /\sqrt{G m_A/a_A^3}$ 
is the orbital period of the inner binary. The constant $k$ depends the inclination of the orbit. For particle orbit inclination close to  $i=90^\circ$, the orbit is librating and $k=k_{\rm L}$, where
\begin{equation}
    k_{\rm L}=\frac{3\sqrt{5}}{4} e_A \sqrt{1+4 e_A^2}
\end{equation}
\citep{Farago2010,Lubow2018},
and for inclination close to  $i=0^\circ$, the orbit is  circulating and $k=k_{\rm C}$, where
\begin{equation}
    k_{\rm C}=\frac{3}{4} \sqrt{1+3 e_A^2 - 4 e_A^4}
\end{equation} 
\citep{Smallwood2019}.
We identify the $k$'s by the subscripts L and C for the precession in a librating or circulating orbit.

For a circular orbit, the coefficient $k$ for arbitrary initial
inclination can be calculated from
\begin{equation}
 k(i_0)
 =
 \frac{3\pi}{8}
 \frac{
 \left[(1-e_{\rm A}^{2})
 (h+4e_{\rm A}^{2})\right]^{1/2}}
 {\mathcal{K}(\kappa^{2})}.
 \label{eq:arbk}
\end{equation}
For the initial nodal phase adopted in our simulations,
$\phi=90^\circ$, the conserved quantity $h$ is
\begin{equation}
 h = \cos^{2}i_0 - 4e_{\rm A}^{2}\sin^{2}i_0,
 \label{eq:h_general}
\end{equation}
where $i_0$ is the initial inclination of the test particle orbit
relative to the inner binary orbital plane, and
\begin{equation}
 \kappa^{2}
 =
 \frac{
 5e_{\rm A}^{2}(1-h)}
 {(1-e_{\rm A}^{2})(h+4e_{\rm A}^{2})}
 =
 \frac{5e_{\rm A}^{2}}
 {1-e_{\rm A}^{2}}
 \tan^{2}i_0.
 \label{eq:kappa_general}
\end{equation}
Here, $\mathcal{K}$ is the elliptic integral
\begin{equation}
\mathcal{K}(\kappa^{2})
=
\begin{cases}
\displaystyle
\int_{0}^{\pi/2}
\frac{d\varphi}
{\sqrt{1-\kappa^{2}\sin^{2}\varphi}},
& \kappa^{2}<1, \\[14pt]
\displaystyle
\int_{0}^{\varphi_{0}}
\frac{d\varphi}
{\sqrt{1-\kappa^{2}\sin^{2}\varphi}},
& \kappa^{2}>1,
\end{cases}
\label{eq:elliptic_integral}
\end{equation}
where
\begin{equation}
 \varphi_{0}
 =
 \sin^{-1}\left(\frac{1}{\kappa}\right).
\end{equation}
The cases $\kappa^{2}<1$ and $\kappa^{2}>1$ correspond to
circulating and librating trajectories, respectively. At the separatrix, $\kappa^{2}=1$, the precession period
diverges \citep{Farago2010}. At $i_0=0^\circ$,
Equation~(\ref{eq:arbk}) reduces to the circulating value
$k_{\rm C}$ given above, while in the limit
$i_0\rightarrow90^\circ$ it reduces to the polar
librating value $k_{\rm L}$.

\subsection{Torque from the outer binary companion}

The torque from the outer binary drives nodal circulation relative to the outer binary angular momentum vector. In addition, if the inclination of the particle orbit relative to the outer binary is large, Kozai-Lidov \citep[KL,][]{Kozai1962,Lidov1962,Zeipel1910} oscillations of the particle may be driven. The inclination of the particle orbit relative to the outer binary and the eccentricity of the particle orbit are exchanged \citep[e.g.][]{Naoz2016}.  The timescale for the particle KL oscillations may be significantly shorter than the timescale for the triple star to undergo KL oscillations.

KL oscillations of the particle can lead to instability since they increase the particle eccentricity and drive close interactions with the inner binary.  The timescale for nodal circulation around the inner binary  is typically much longer than the polar libration timescale. Therefore, KL oscillations of the particle may be more likely for triple star systems with a large mutual inclination between the inner and outer binary.

For a particle orbiting at radius $r$, timescale for KL oscillations is approximated by
{\begin{equation}
    t_{\rm KL}= \frac{m_{AB}}{m_B} \frac{P_{AB}^2}{{P_A}} (1-e_{AB}^2) \left(\frac{a_{AB}}{r}\right)^{3/2}\,,
\end{equation}
\citep[e.g.][]{Kiseleva1998},
where the orbital period of the outer binary is $P_{AB}=2\pi/\sqrt{G m_{AB}/a_{B}^3}$. 
Note that the timescales are only approximate but they fit to numerical experiments \citep{Martin2022}.

\subsection{Critical radius where the torques balance}

The nodal precession timescale increases with the test particle radius, whereas the KL timescale falls off.  The radius where these two are equal represents the transition from the dominant torque being the inner binary to the outer binary.
Using our two approximate values we find a critical radius where the two time scales are equal
\begin{equation} 
    \frac{R_{\rm crit}}{a_A} = \bigg[k\frac{m_{Aa}m_{Ab}}{m_A m_B} \bigg(\frac{a_{AB}}{a_A}\bigg)^3 \\  (1-e_{AB}^2) 
    \bigg]^{1/5}    .
\label{eq:rc}
\end{equation}
Outside this critical radius, $r>R_{\rm crit}$, the KL oscillations can make the test particles unstable.
For the librating region,  substituting $k=k_L$, we define a critical radius $R_{\rm crit}=R_{\rm L}$ as the maximum extent of the librating region, and for the circulating region substituting $k=k_{\rm C}$  we find $R_{\rm crit}=R_{\rm C}$ as the maximum extent of the circulating region.
Strictly speaking, $R_{\rm L}$ is appropriate for polar librating test particle orbits around a coplanar triple star system and $R_{\rm C}$ for coplanar orbits in a highly misaligned system.  However, we will see that the $R_{\rm L}$ provides a good guide to the librating region for all mutual misalignments. We can also calculate a critical radius for an arbitrary inclination by using our $k(i)$ defined above in equation~(\ref{eq:arbk}).

\subsection{Stationary states} 

There are two stationary states for particle orbits that do not undergo nodal precession. First, the particle orbit may be coplanar, or retrograde coplanar to the inner binary. Second, the particle may be in a polar orientation if the inner binary is eccentric. In the absence of the outer companion, the polar stationary state is aligned to the eccentricity vector of the inner binary.

The outer companion drives apsidal precession of the inner binary that leads to the stationary inclination changing with the particle orbital radius \citep[e.g.][]{Lepp2023}. The stationary inclination is where the particle nodal precession rate balances the inner binary apsidal precession rate.  In this case there is additional nodal precession on the test particle from the outer binary and we find that the stationary state stays very close to $90^\circ$ value, this is similar to that seen in \cite{Verrier2009} 
While this effect is present we find that the dynamics is dominated by the quadrapole of the outer binary.

\section{Simulation set-up}
\label{sec:sim}

\begin{figure*} 
\begin{centering} 
\includegraphics[width=8cm]{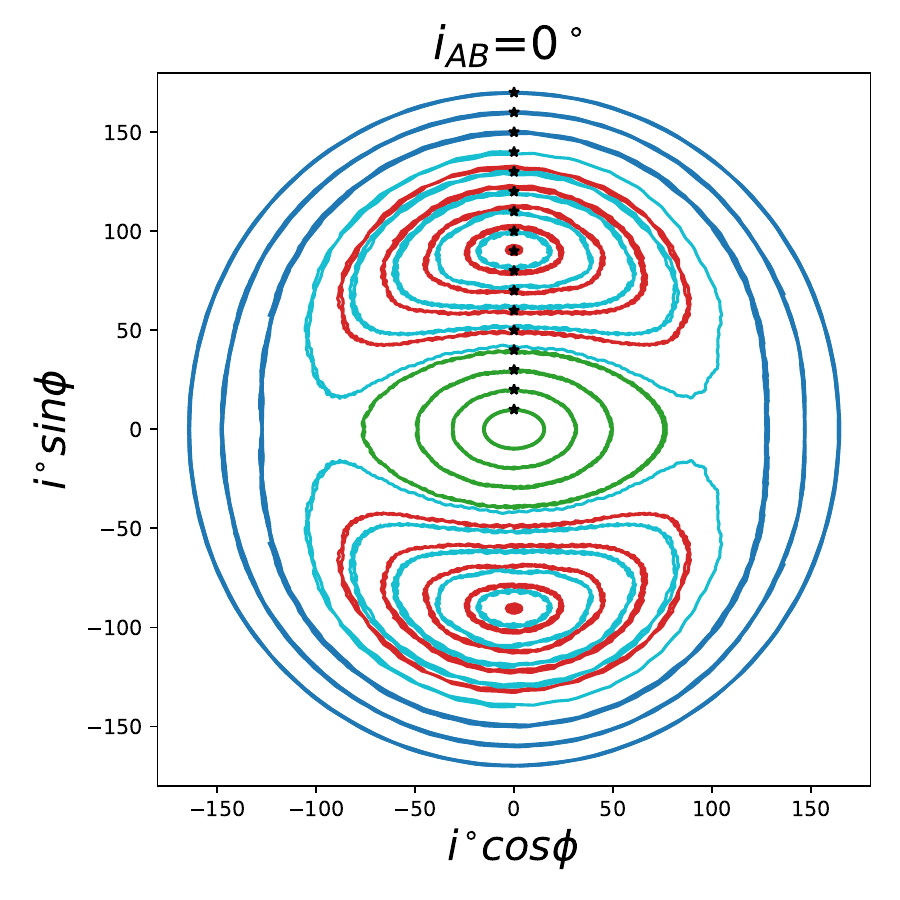}
\includegraphics[width=8cm]{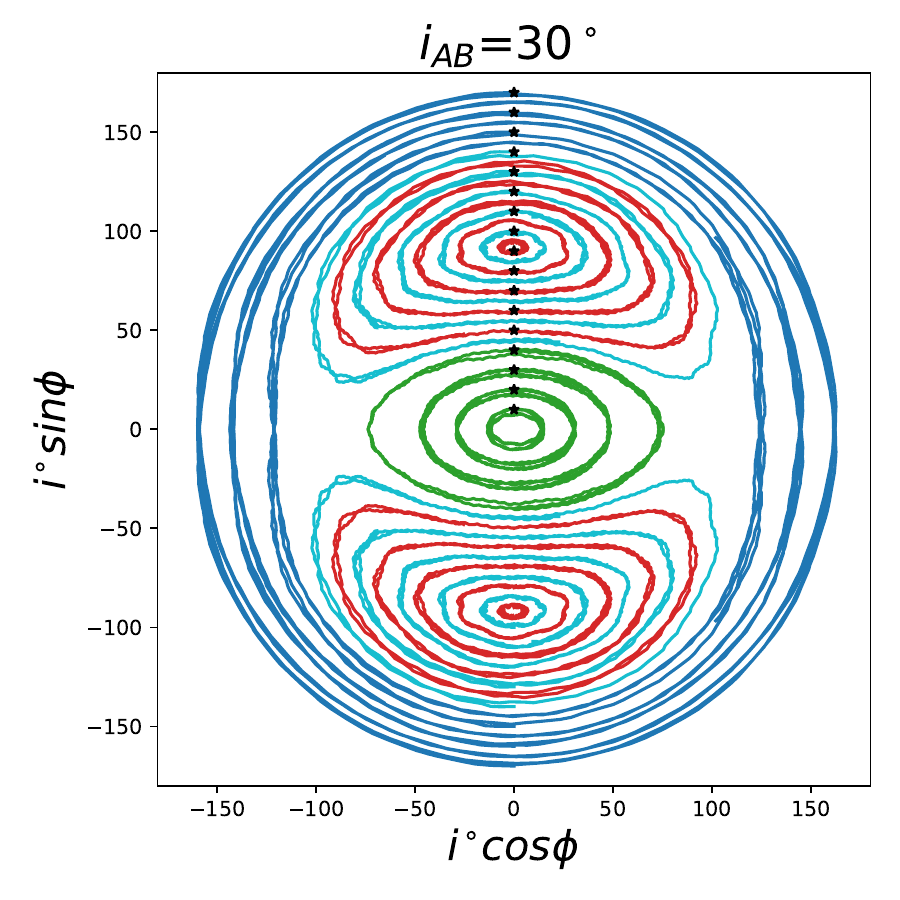}
\includegraphics[width=8cm]{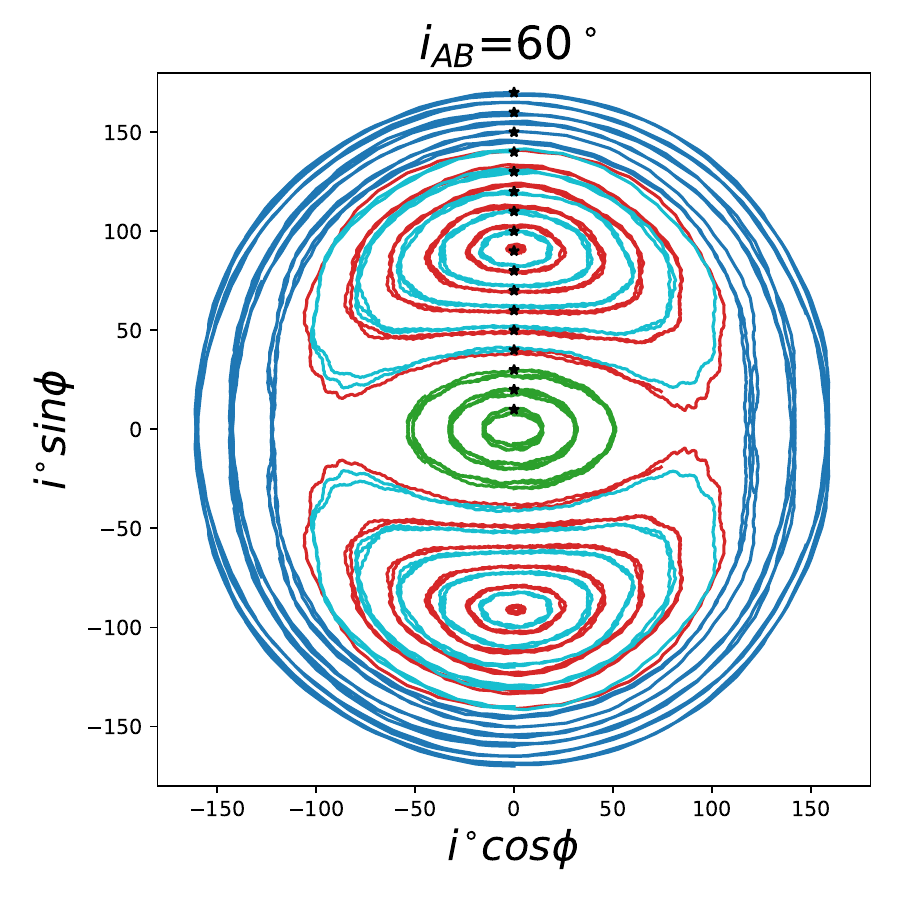}
\includegraphics[width=8cm]{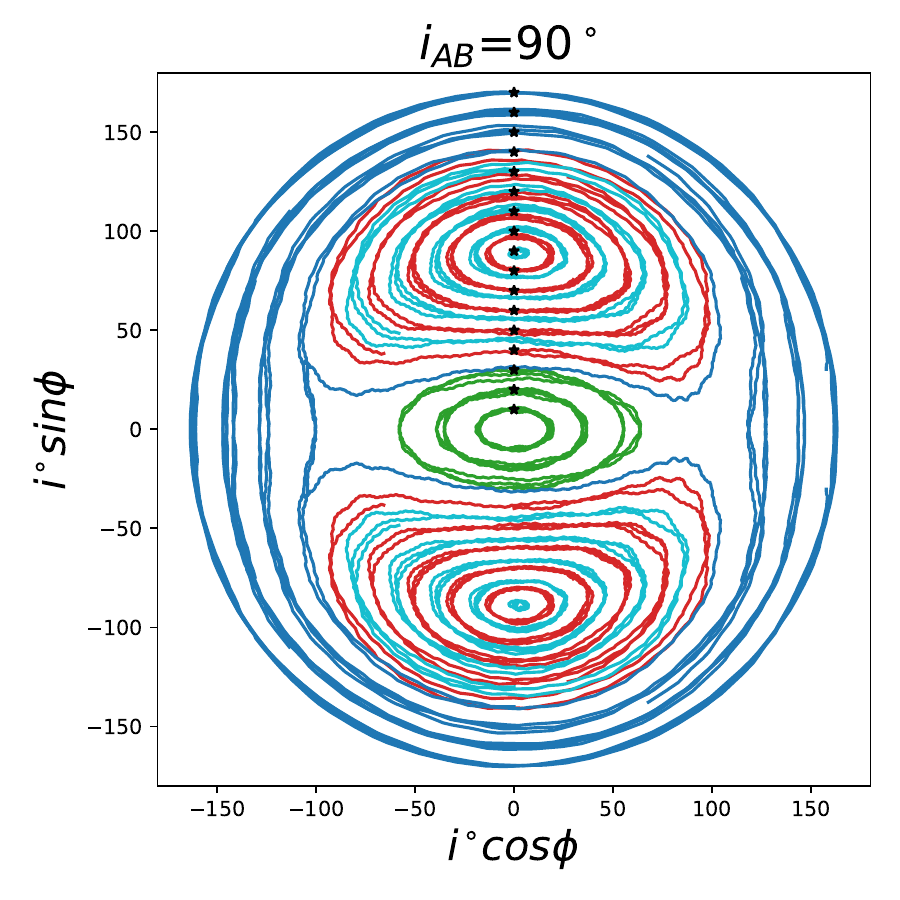}
\vspace{0.2cm}
    \caption{The phase diagram  $(i\cos\phi,i\sin\phi)$ for circumbinary test particle orbits in the frame of the  inner binary for four mutual inclinations to the outer companion ($i_{AB}= 0^\circ, 30^\circ, 60^\circ \text{ and } 90^\circ$) at a test particle radii ($r= 4\, a_{A}$). The initial inclinations of the test particle relative to the outer binary are  $i=10^\circ \, \text{through } \,170^\circ$ in steps of $10^\circ$, with the starting point marked by a black star.  The orbits are colored green for prograde circulating orbits, red (cyan) for librating orbits that have an initial inclination lower (higher) than the stationary inclination, and blue for retrograde circulating orbits. Each test particle orbit is plotted twice,  normal and reflected about the x-axis, in order to show the librating region in the lower half. }
\label{phasemap1} 
\end{centering} 
\end{figure*}

The $n$-body simulations in this paper use the {\sc rebound} $n$-body code \citep{rebound}. The simulations for the parameter study are integrated using WHFast, a symplectic Wisdom-Holman integrator \citep{reboundwhfast,wh}. 
 WHFast was used because it was much faster than IAS15. However, for the standard model  parameters we also used IAS15, a 15th order Gauss-Radau integrator \citep{reboundias15}. For example, in reproducing a stability map with our standard parameters (the upper right panel of Fig.~\ref{stabilitymap1}, see Section~\ref{sec:results}), the number of
stable initial conditions was 19513 with IAS15, and 19506 with WHFast.  Only a few pixels in the panel were changed and so we choose to use WHFast to reduce the computation time.

We examine test particle orbits about the inner binary described in Section~\ref{sec:standard}.  The test particles begin in circular orbits at semimajor axis  $r$ and at inclination $i$ relative to the inner binary.  The inclination is defined as 
\begin{equation}
i = \cos^{-1}(\hat{\bm{l}}_{A}\cdot \hat{\bm{l}}_{\rm t})\,,
\end{equation}
where $\hat{\bm{l}}_{A}$ and $\hat{\bm{l}}_{\rm t}$ are the unit vectors in the direction of the angular momentum of the inner binary and the test particle respectively.
We also define a nodal  phase angle of the test particle as the angle measured relative to 
the eccentricity vector of the inner binary and is 
given by 
\begin{equation}
        \phi = \tan^{-1}\left(\frac{\hat{\bm{l}}_{\rm t}\cdot (\hat{\bm{l}}_{A}\times 
    \hat{\bm{e}}_{A})}{\hat{\bm{l}}_{\rm t}\cdot \hat{\bm{e}}_{A}}\right) + 90^\circ 
\end{equation}
\citep{Chen2019,Chen2020e}, where  $\hat{\bm{e}}_{A}$ is the  unit eccentricity vector of the inner binary. We start the test particle orbits with an initial phase angle of $\phi=90^\circ$. For our standard setup this gives us the all possible test particle orbits, because both circulating and librating orbits pass through $\phi=90^\circ$.  

Fig.~\ref{phasemap1} shows phase diagrams in the $(i\cos\phi,i\sin\phi)$ plane for some circumbinary test particle orbits.  The triple star system has our standard parameters for four different initial mutual inclinations $i_{AB}=0^\circ$, $30^\circ$, $60^\circ$ and  $90^\circ$.  The  test particles begin in circular orbits around the inner binary at a radius of $r=4 \, a_{A}$.  This is an orbit close to the inner binary and so the torque from the inner binary dominates the dynamics.   The initial test particle inclinations relative to the inner binary are  $i=10^\circ$ through  $170^\circ$ in steps of $10^\circ$.

The test particle orbits are colored by their orbit type. 
Green shows circulating orbits.  Red shows librating orbits that start with an inclination that is lower than the polar stationary inclination.   Cyan shows orbits which are librating orbits and whose initial inclination is higher than the polar stationary inclination.  Finally,  blue shows orbits which are retrograde circulating orbits.  
In the lower right panel, the orbit which began at an initial inclination of $140^\circ$ is sometimes circulating and sometimes librating,  this can occur because of influences of the third body,  it is colored as a circulating orbit.  In our classification scheme, inclinations below this are classed as circulating orbits.

The cleanest phase map is for the coplanar triple star case ($i_{AB}=0^\circ$) as perturbations to the inner binary are limited to the prograde precession caused by the outer companion.  With higher mutual inclinations, the interactions from the companion and the inner binary make the track of the orbital parameters more complicated but at this radius there are still clear circulating, librating and retrograde circulating regions.

\begin{figure*} 
\begin{centering}

\hspace{2cm}$\Omega_{AB}=0^\circ$  \hspace{5cm} Standard Parameters($\Omega_{AB}=90^\circ$)

\vspace{-12pt}

\includegraphics[width=8cm]{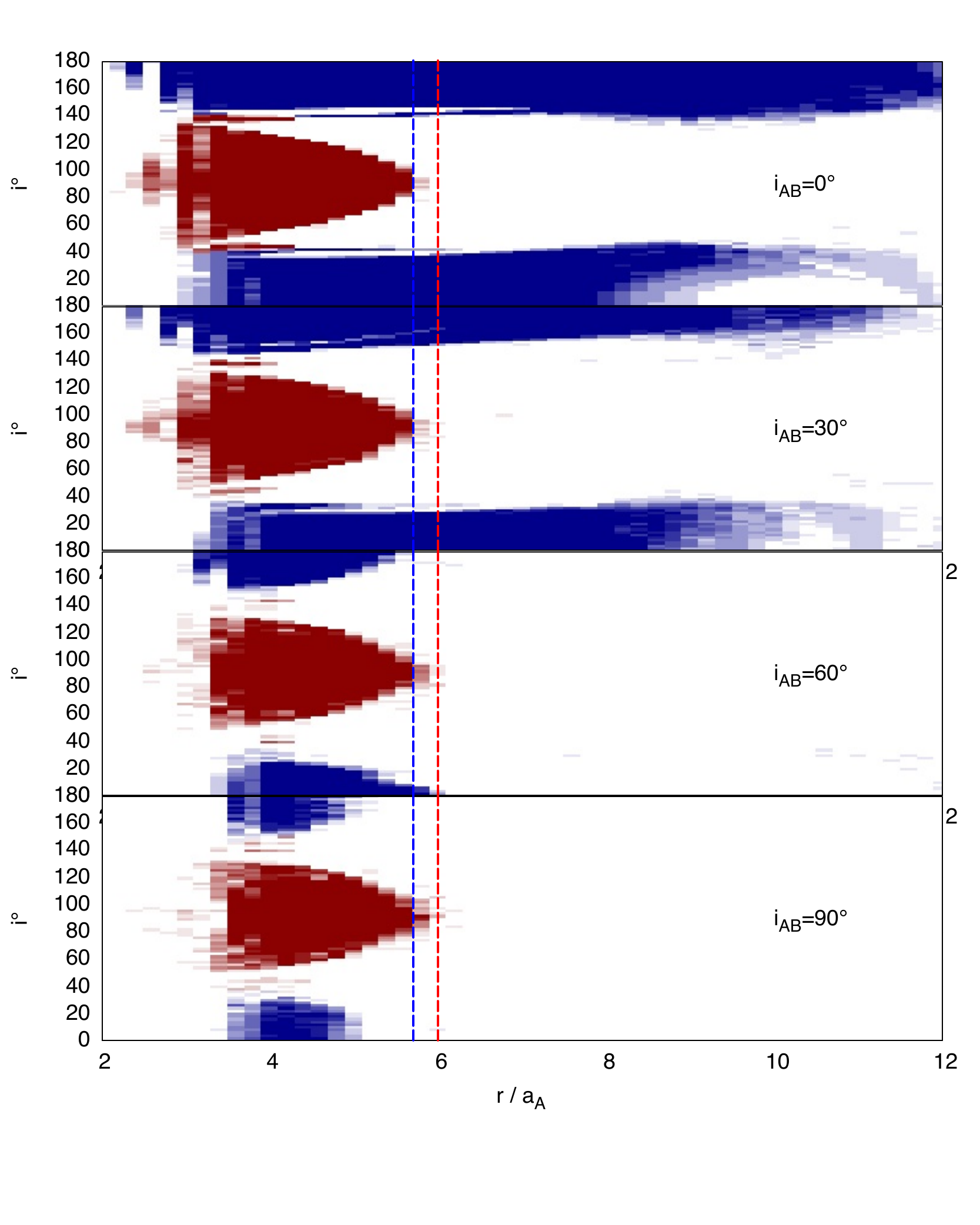}
\includegraphics[width=8cm]{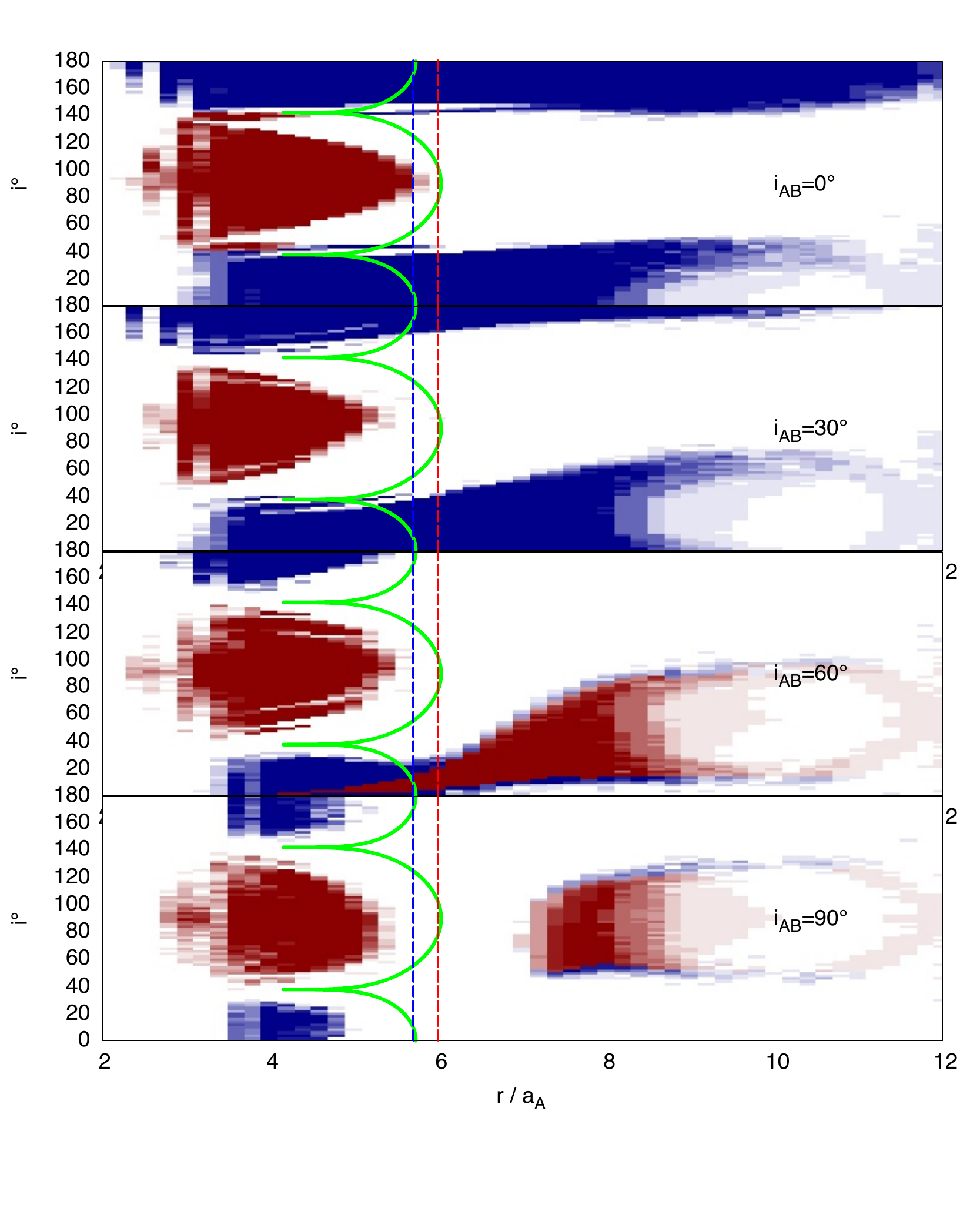}

\vspace{-22pt}

$\Omega_{AB}=180^\circ$  \hspace{7cm} $\Omega_{AB}=-90^\circ$

\vspace{-12pt}

\includegraphics[width=8cm]{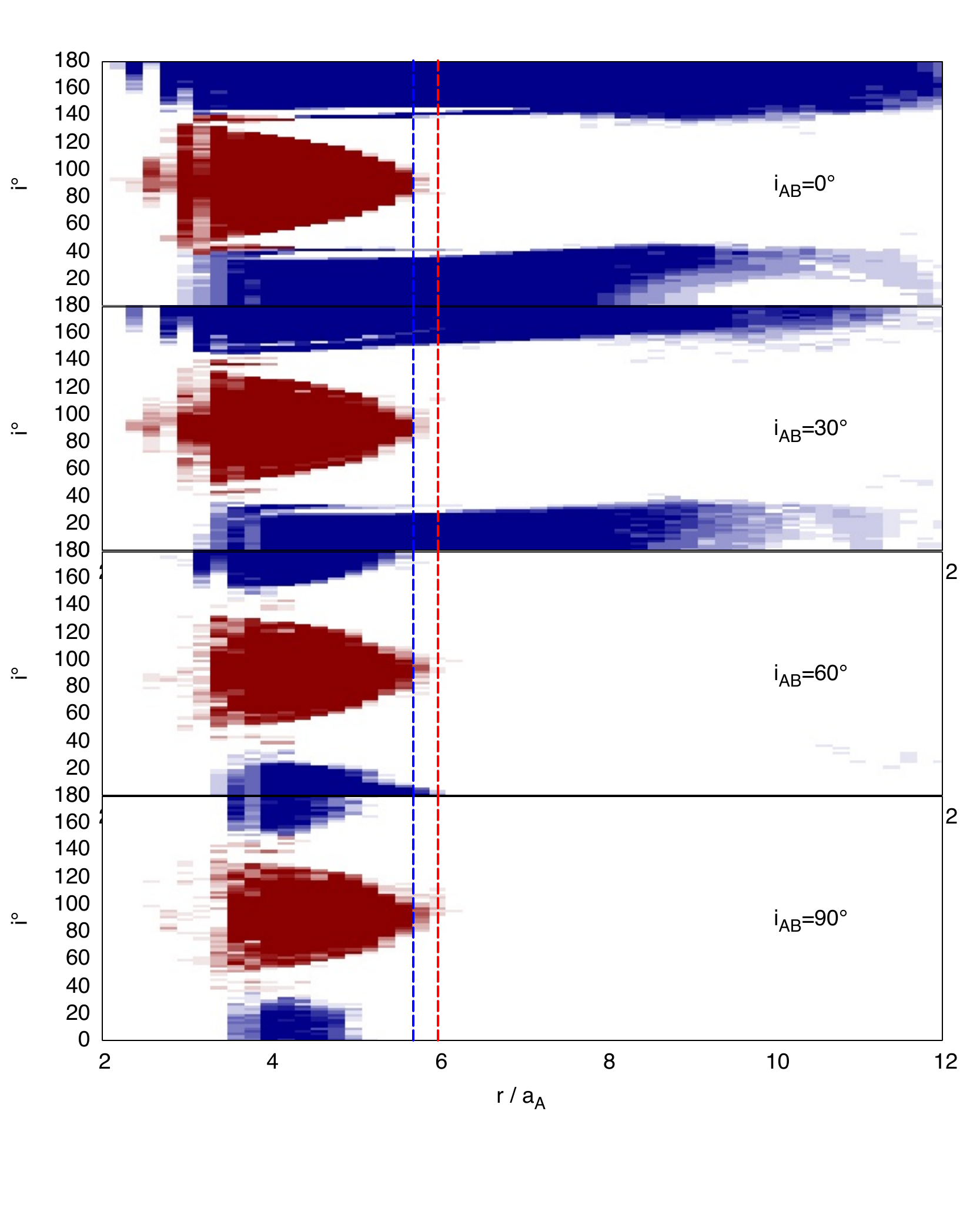}
\includegraphics[width=8cm]{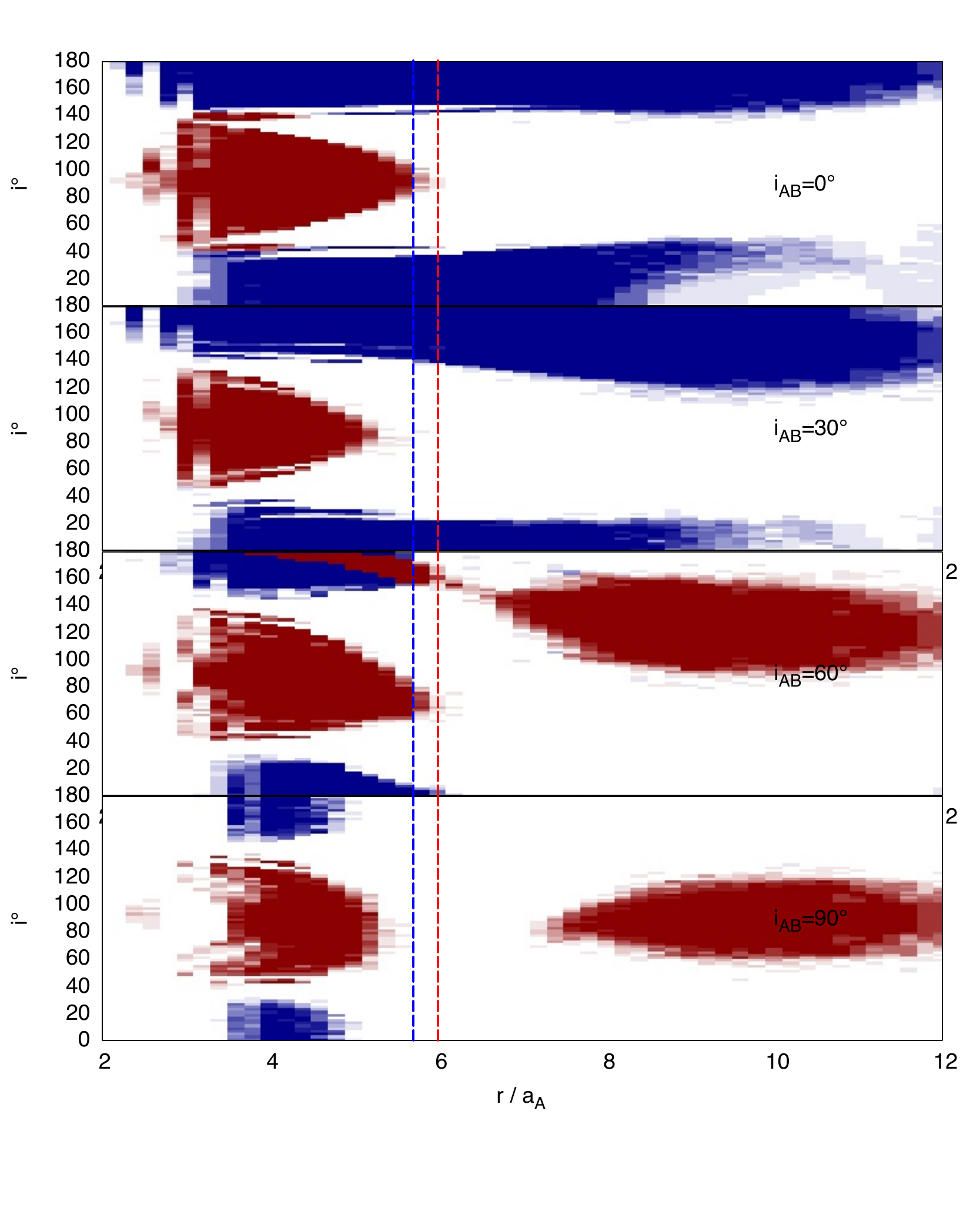}
\vspace{0.2cm}

\vspace{-22pt}

\caption{Stability of test particles orbiting around a binary ($e_{\rm A}=0.5$) with a circular orbit outer stellar companion ($m_{Aa}=m_{Ab} = m_{B}/2$) with semi-major axis ratio  $a_{AB}=30\,a_A$.  There are four sets of four panels which are calculated with a longitude of the ascending node for the companion of $\Omega_{AB}=0^\circ$ (top left panel), $90^\circ$ (top right panel), $180^\circ$ (bottom left panel) and $-90^\circ$ (bottom right panel). In each panel, the inclinations for the companion are $i_{AB}=0^\circ$ (top row), $30^\circ$ (second row), $60^\circ$ (third row) and $\,90^\circ$ (fourth row). 
The particles begin with inclination $i$ relative to the inner binary orbit with semi-major axis $r$. Red and blue regions show where the particles are stable for a time of at least $t=50,000\,P_A$; this is about twice the KL oscillation timescale of the triple star. White regions are unstable. The vertical red and blue lines show $R_{\rm L}=6.0\, a_A$ and $R_{\rm C}=5.7\, a_A$, respectively. The green line in the upper right panel shows the critical radius for an arbitrary inclination calculated with equation~(\ref{eq:arbk}).  }
\label{stabilitymap1} 
\end{centering} 
\end{figure*}

\section{Stability maps}
\label{sec:results}

We now examine the stability of test particles orbiting a binary star with an outer stellar companion. For each set of triple star parameters, we consider a range of initial orbital radii and inclinations for the particles.  The initial orbital radius is in the range $r=2-12\,a_A$ and is stepped by $0.1\,\rm  au$.  The initial inclination relative to the inner binary orbit is in the range $i=1^\circ-179^\circ$ and is stepped by $2^\circ$. The initial longitude of ascending node is $\phi=90^\circ$.
For the companion we examine four different initial values of $\Omega_{AB}$. The initial mutual inclination between the test particle orbit
and the outer binary orbit, $i_{\rm Bt}$, depends upon the inclination and longitude of ascending node of both the particle and the outer binary according to 
\begin{equation}
\cos i_{\rm Bt} =
\cos i \cos i_{\rm AB}
+\sin i \sin i_{\rm AB}
\cos(\phi-\Omega_{\rm AB}).
\end{equation}
For each initial combination of orbital radius and inclination, we run 10 different simulations starting with different initial true anomalies equally spaced from $0-360^\circ$.  Each simulation is run for a time of 50,000$\,P_{A}$.

An orbit is labeled as unstable if the eccentricity of the test particle orbit becomes larger than 1, if the semimajor axis becomes larger than $\,a_{AB}$,  or if the semimajor axis becomes less than $a_A$ \citep[see for example][]{Quarles2018,Chen2020}. If the orbit is stable then it is checked to see if it is librating or circulating with respect to the inner binary.  If the orbit stays in the upper half of the phase diagram then it is considered a librating orbit. All the others are circulating.  If at least one of the orbits
for a given initial radius and inclination
is librating then the pixel is considered librating (note it is rare to have both in the same pixel).
The hue denotes the orbit classification, with red indicating librating orbits and blue indicating circulating orbits. The color intensity indicates the number of stable simulations, ranging from white (0 stable orbits) to saturated red or blue (10 stable orbits).

The simulation time of $50,000\,P_A$ was chosen to be consistent with \cite{Chen2020}. It is equal to about 400$\, P_{AB}$ and about 25 KL cycles of test particle and 2 KL cycles of triple star. We examined the dependence on the integration
time by repeating the standard model for $2$, $3$, and
$10$ times longer. For example, for the top right panel of Fig.~\ref{stabilitymap1}, the number of stable initial
conditions decreased from 19506 to 19047, 18824, and
18197, respectively. Thus, increasing the integration
time by an order of magnitude reduces the number of
stable orbits by only about $7\%$, while the overall
structure of the stability maps and the locations of the
principal stability boundaries remain essentially
unchanged.  Therefore, all of the simulations we present have an end time of $50,000\,P_A$.

The simulations are scale free with respect to separation and mass, and so the time represented by the 50,000\,$P_A$ depends on the scales.  If we take $m_{Aa} = m_{Ab} = 1\, M_\odot$  and the separation of $a_A= 1\,$au then the period is about $P_A=0.707\,$yr and $50,000\,P_A$  corresponds to a time of about $3.5\times 10^4\,$yr. Our longest simulation time is 10 times longer, giving times comparable but slightly shorter than a protoplanetary disk lifetime of about $10^6\,$yr.

\subsection{Standard parameters}
\label{sec:standard_results}

We first consider our standard model parameters with
$\Omega_{\rm AB}=90^\circ$. The upper-right panel of
Fig.~\ref{stabilitymap1} shows the stability maps as the mutual
inclination between the inner and outer binaries is varied through
$i_{\rm AB}=0^\circ$, $30^\circ$, $60^\circ$, and $90^\circ$.
With this choice of $\Omega_{\rm AB}$, the test particle orbits and
the outer companion are initially tilted in the same direction.

The stability maps contain three main regions. Close to the inner
binary there is a stable prograde circulating region at low particle
inclination, a stable retrograde circulating region at high
inclination, and an island of stable librating orbits centered near
polar inclination. The inner stability boundary is at approximately
$3\,a_{\rm A}$, similar to that found by \citet{Chen2020}.
There is a tendency for this boundary to move outward as
$i_{\rm AB}$ increases. This may result from KL oscillations of the
inner binary driven by the companion, which temporarily increase
the inner-binary eccentricity; binaries with larger eccentricity
have a larger inner stability boundary \citep[see][]{Holman1999}.

The outer extent of the librating region is relatively insensitive
to the mutual inclination of the two binaries. The most stable
librating orbits remain close to $i=90^\circ$, and their outer
stability boundary is in good agreement with $R_{\rm L}$.
The radial extent of the circulating regions, however, is strongly
dependent on $i_{\rm AB}$. At large mutual inclinations, particles
that are initially close to coplanar or retrograde coplanar with
the inner binary can undergo KL oscillations driven by the outer
companion. The circulating stable regions are therefore substantially
reduced, with their outer boundary approaching $R_{\rm C}$ for
large $i_{\rm AB}$.

A stable region also extends to larger radii along inclinations close
to alignment with the outer binary orbit. An unstable region is
centered near $r\simeq10\,a_{\rm A}$ and at approximately the
inclination of the companion. This region lies roughly two-thirds
of the way to the center of mass of the three-star system, located
at $r=15\,a_{\rm A}$, where the potential due to the companion is
approximately half that due to the inner binary. The instability
begins at about $r\simeq8\,a_{\rm A}$, consistent with the outer
stability limit expected for an orbit about one component of a
binary \citep{Holman1999}. Although some orbits beyond this region
are stable according to our adopted criterion, their trajectories
need not remain simple in the frame of the inner binary.

For $i_{\rm AB}=60^\circ$, the extended stable region is circulating
at small particle radii but becomes librating at larger radii. As
the particle radius increases, the quadrupole perturbation from the
companion becomes increasingly important and shifts the center of
the circulation away from zero inclination. The resulting trajectories
are classified as librating according to our criterion that
$i\sin\phi$ does not change sign. This behavior is analogous to the
transition across the Laplace surface for an oblate planet orbiting
a star \citep[e.g.][]{Tremaine2009,lubow2025}, and is discussed
further in Section~\ref{sec:discuss}.

Also shown on the stability maps are red and blue vertical lines show the critical radii
$R_{\rm L}$ and $R_{\rm C}$, respectively, from
Equation~(\ref{eq:rc}). For our standard parameters these are
$R_{\rm L}=6.0\,a_{\rm A}$ and $R_{\rm C}=5.7\,a_{\rm A}$.
For our standard model we also plot a green line which shows the critical radius as a function of initial
test particle inclination calculated using $k(i)$ from
Equation~(\ref{eq:arbk}).

The critical radii $R_{\rm C}$ and $R_{\rm L}$ apply where KL
oscillations are active. For $i_{\rm AB}=0^\circ$ and
$30^\circ$, $R_{\rm L}$ describes the outer extent of the
librating region, while for $i_{\rm AB}=60^\circ$ and
$90^\circ$, $R_{\rm C}$ approximately describes the outer extent
of the circulating regions. However, the nodal precession timescale
of a test particle about the outer-binary angular momentum vector is
similar to the KL timescale and can also affect the stability of
librating orbits. Polar librating orbits require the test particle angular
momentum vector to remain close to the inner-binary eccentricity
vector. Nodal precession that drives the particle away from this
alignment can disrupt the polar configuration. Consequently,
$R_{\rm L}$ provides a useful estimate of the extent of the
librating region for all values of $i_{\rm AB}$.

For circulating orbits, by contrast, nodal precession causes the
test particle angular momentum vector to precess about the inner-binary
angular momentum vector without necessarily destabilizing the orbit.
Thus, $R_{\rm C}$ is relevant primarily where KL oscillations
operate, namely for circulating orbits at large $i_{\rm AB}$.
For the same reason, the inclination-dependent critical radius shown
by the green line provides a good estimate of the circulating
stability boundary primarily at the larger mutual inclinations.

The stable regions, particularly those within the critical radii,
undergo only modest eccentricity excitation. The eccentricities of
test particles in these regions generally remain below 0.1.

\subsection{Effect of varying the longitude of the ascending node}
\label{sec:omega}

Figure~\ref{stabilitymap1} also shows the stability maps for
$\Omega_{\rm AB}=0^\circ$, $180^\circ$, and $-90^\circ$,
allowing us to examine the dependence on the initial longitude of
ascending node of the outer binary. The critical radii calculated
above are independent of $\Omega_{\rm AB}$ and are therefore the
same in all four panels. We center our discussion in relation to $R_L$ and $R_C$ and so for clarity, the inclination-dependent
critical radius shown by the green line in the
$\Omega_{\rm AB}=90^\circ$ panel is omitted from the other panels.

The $\Omega_{\rm AB}=0^\circ$ and $180^\circ$ stability maps are
quite similar. For $\Omega_{\rm AB}=0^\circ$, the
$i_{\rm AB}=0^\circ$ case is coplanar, while at
$i_{\rm AB}=90^\circ$ the two binary orbital planes are
perpendicular. As in the standard case, the maps contain stable
prograde and retrograde circulating regions and an island of stable
librating orbits near polar inclination. The librating region and
its outer boundary change relatively little with
$\Omega_{\rm AB}$.

For $i_{\rm AB}=0^\circ$ and $30^\circ$ in the
$\Omega_{\rm AB}=0^\circ$ case, there is an unstable region near
$r\simeq10\,a_{\rm A}$ for test particle orbits close to coplanar with
the inner binary. The instability begins near $r\simeq8\,a_{\rm A}$,
consistent with the outer stability limit obtained by treating the
inner binary as a point mass in a binary system
\citep[see][]{Holman1999}.

The $\Omega_{\rm AB}=-90^\circ$ stability map has similarities to
the $\Omega_{\rm AB}=90^\circ$ standard case, including an extended
stable region associated with test particle orbits close to the outer
binary plane. However, the instability centered near
$r\simeq10\,a_{\rm A}$ in the $\Omega_{\rm AB}=90^\circ$ case
is absent. The test particles in this extended region are retrograde
relative to the outer binary orbit, and retrograde coplanar orbits
are more stable than their prograde counterparts
\citep[e.g.][]{Morais2012,Overton2024,Overton2025}.

Thus, varying $\Omega_{\rm AB}$ modifies some of the detailed
structure at large test particle radii, particularly for orbits associated
with the outer binary plane, but has relatively little effect on the
inner librating stability region that is the primary focus of this
work. We therefore adopt $\Omega_{\rm AB}=90^\circ$ for the
remainder of the paper while varying the other binary parameters.
\begin{figure*} 
\begin{centering} 
\vspace{0pt}
a)\,\,$a_{AB}=15\,a_A$
\hspace{4cm} b)\,\, $a_{AB}=60\,a_A$
 
\includegraphics[width=8cm]{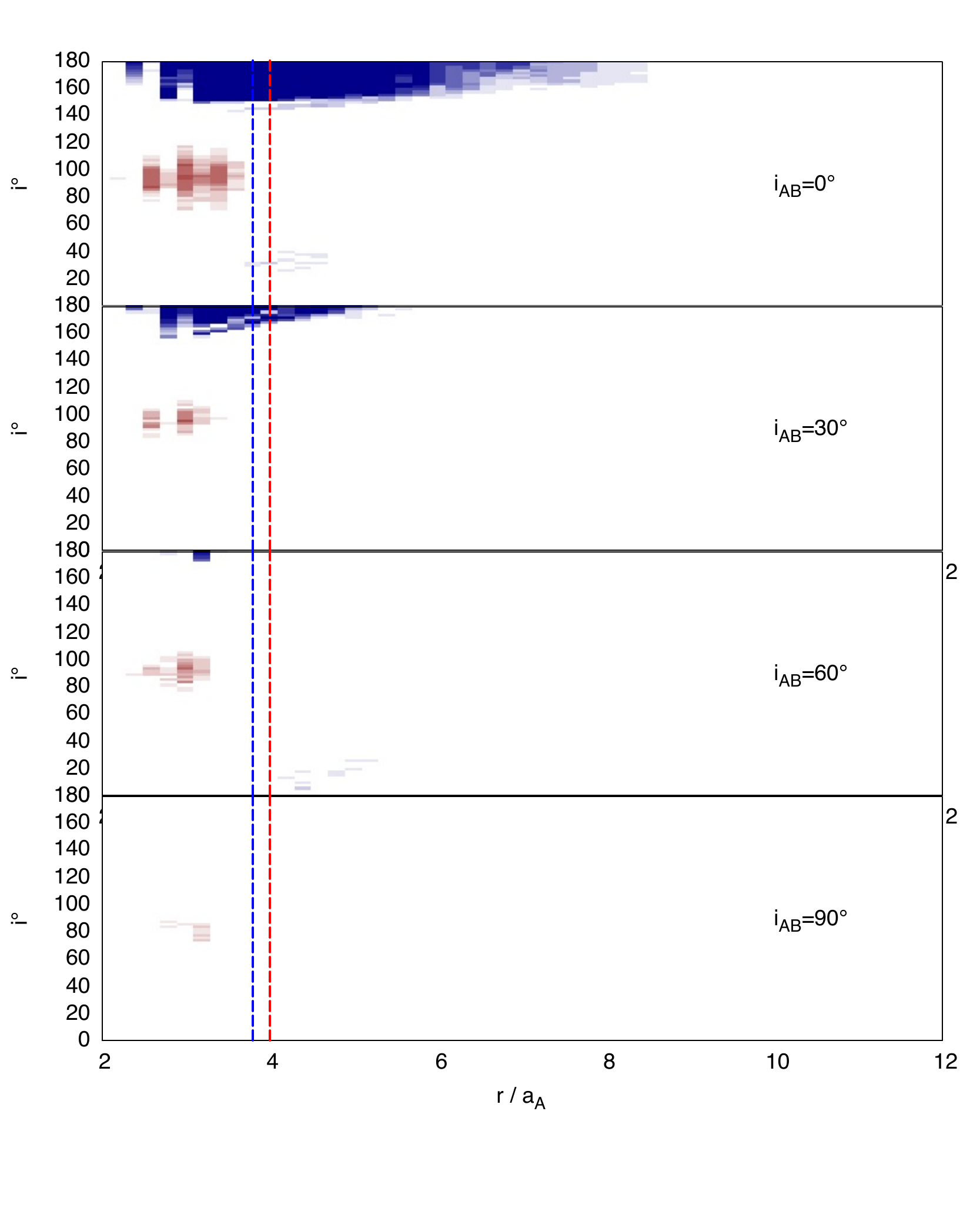}
\includegraphics[width=8cm]{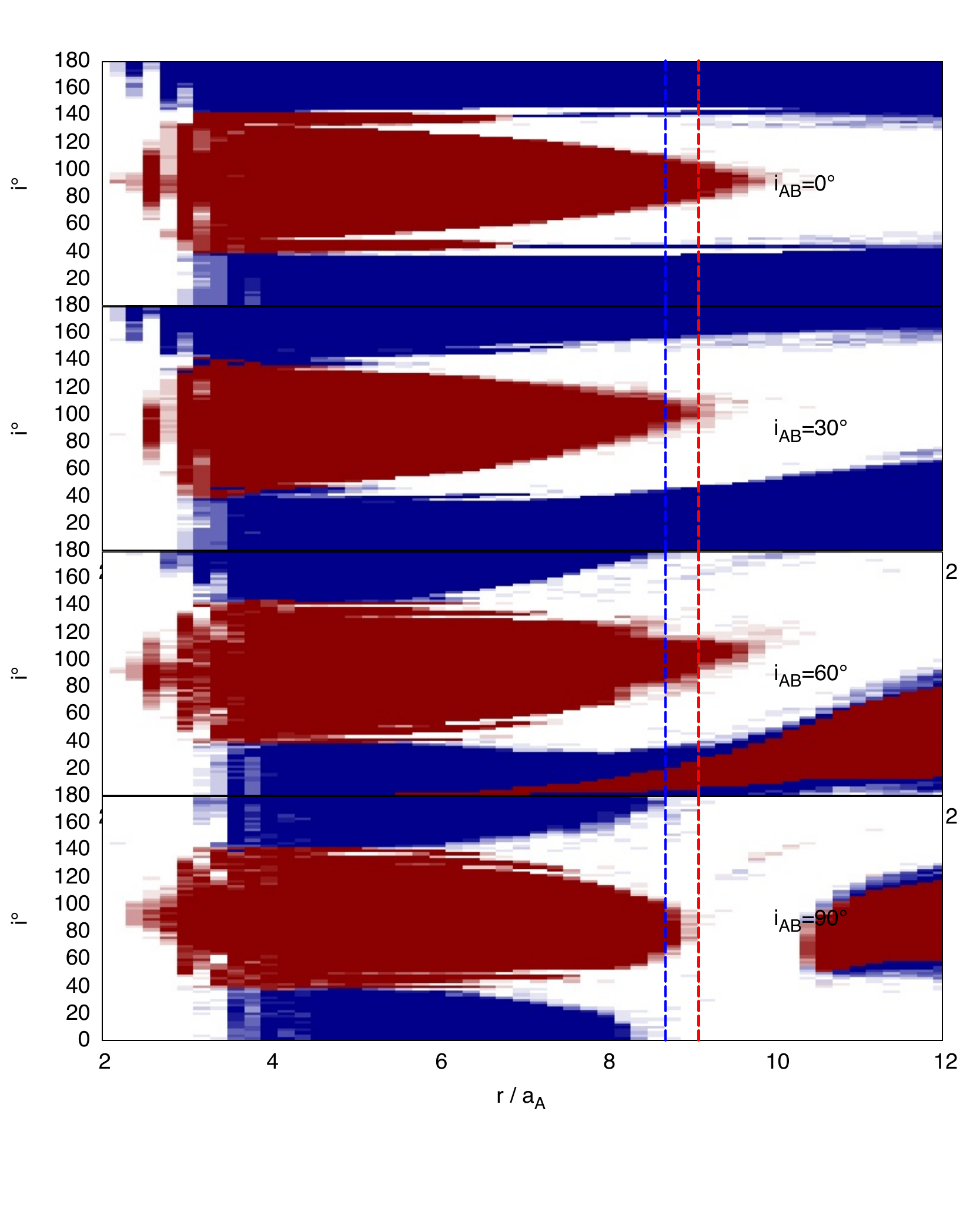}

\vspace{-20pt}

c)\,\,$m_B=0.5\,m_A $
\hspace{4cm} d)\,\, $m_B=2\,m_A$ 

\includegraphics[width=8cm]{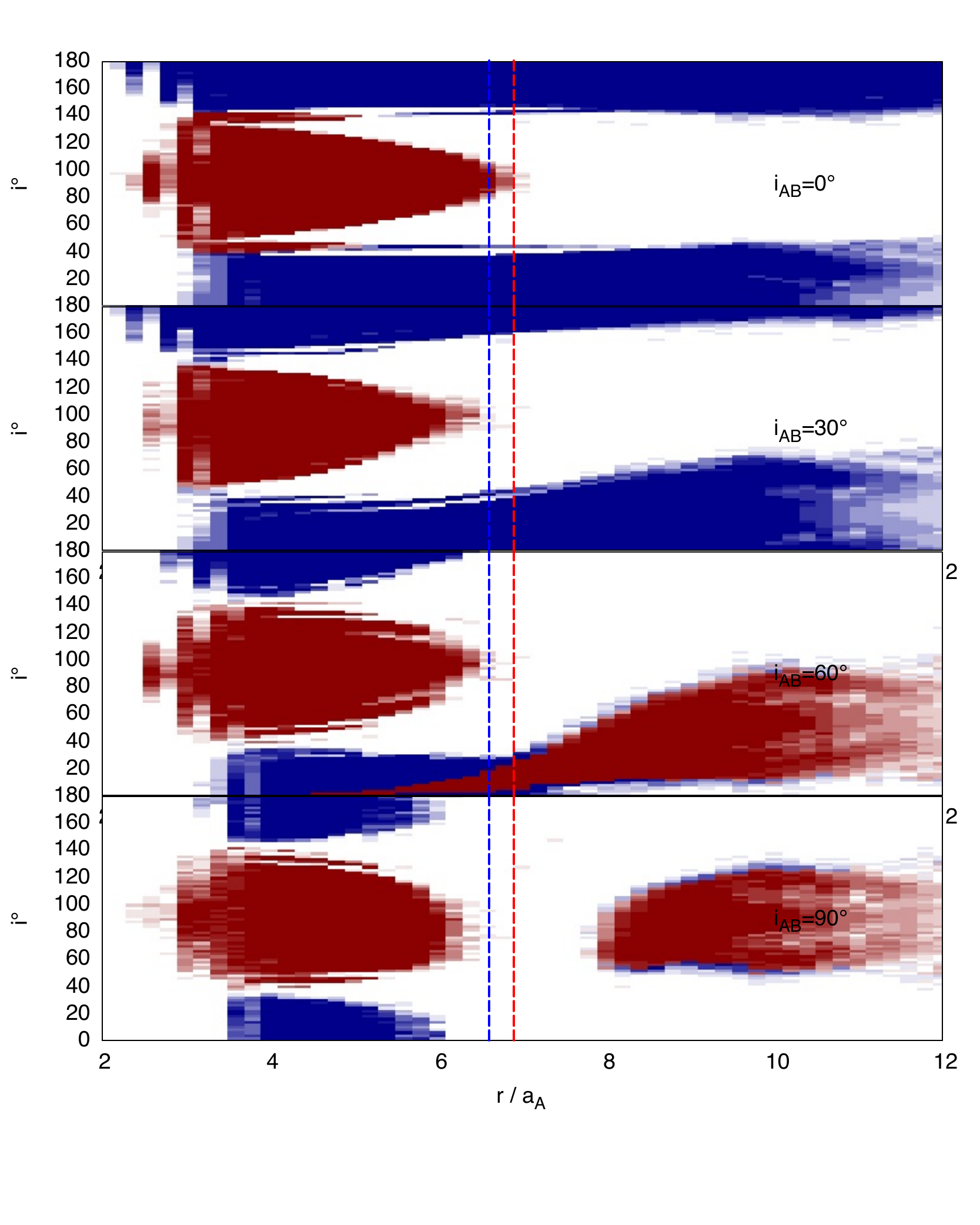}
\includegraphics[width=8cm]{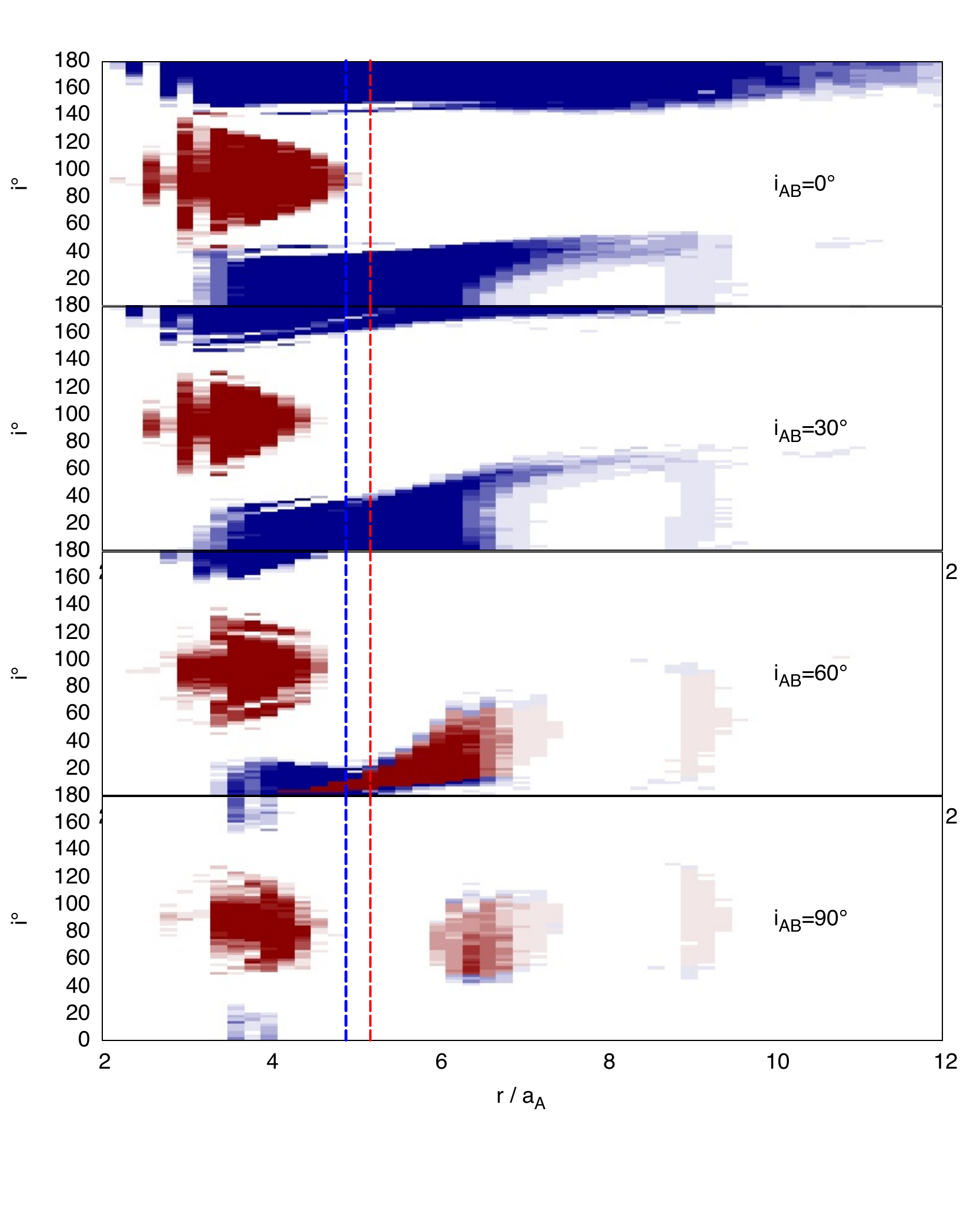}
\vspace{0.2cm}
\caption{Each panel is the same as the upper right panel in Fig.~\ref{stabilitymap1} with $\Omega_{AB}=90^\circ$ and the standard triple star parameters except one parameter is changed in each. The upper left plot has $a_{AB}=15\,a_A$, the upper right has $a_{AB}=60\,a_A$, the lower left has $m_B=0.5\,m_A$ and the lower right has $m_B=2\,m_A$.  }
\label{stab:p1} 
\end{centering} 
\end{figure*}

\begin{figure*}[p]
\begin{centering} 
\vspace{0pt}
a)\,\,$e_A=0.2$
\hspace{5cm} b)\,\,$e_A=0.8$

\includegraphics[width=8cm]{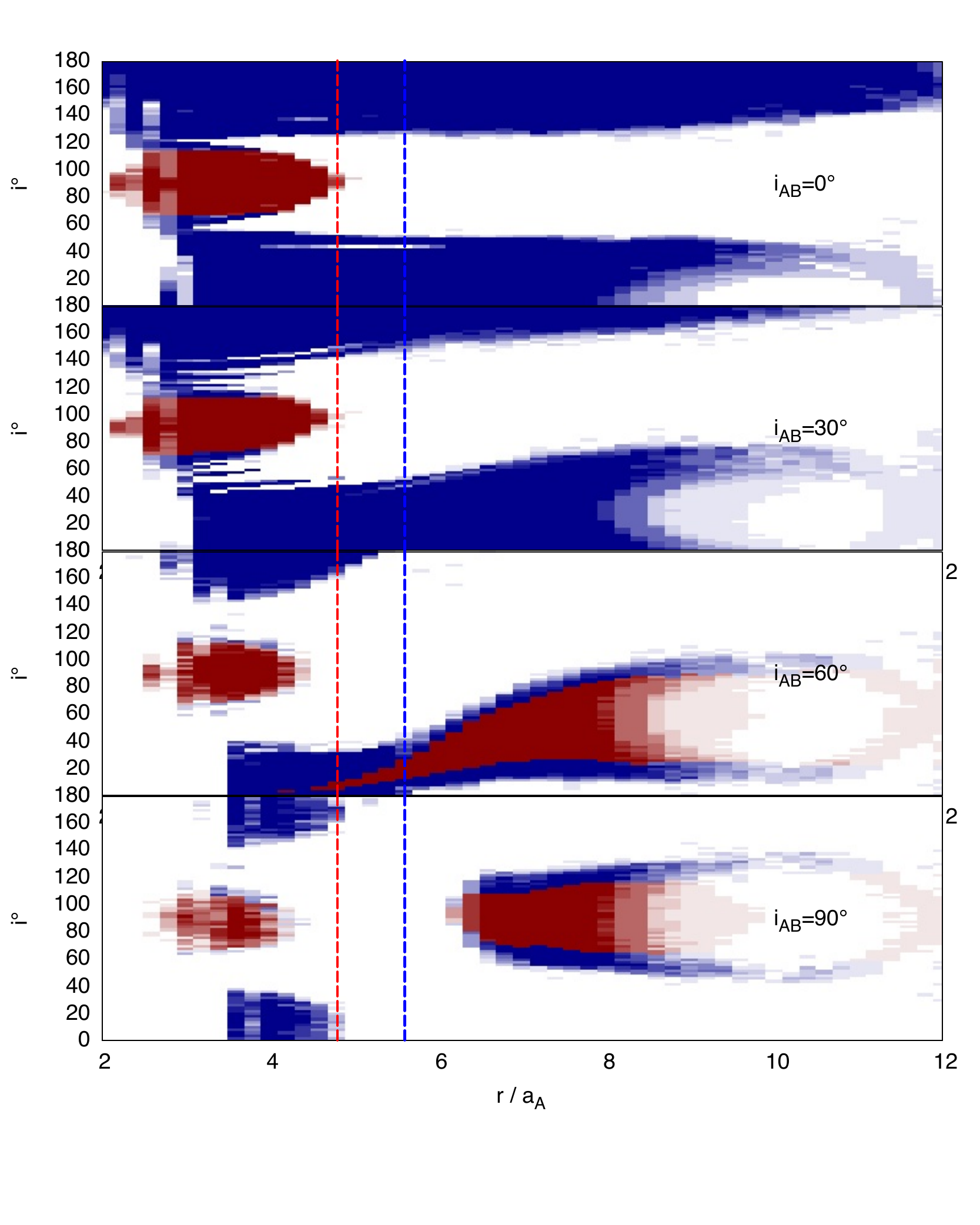}
\includegraphics[width=8cm]{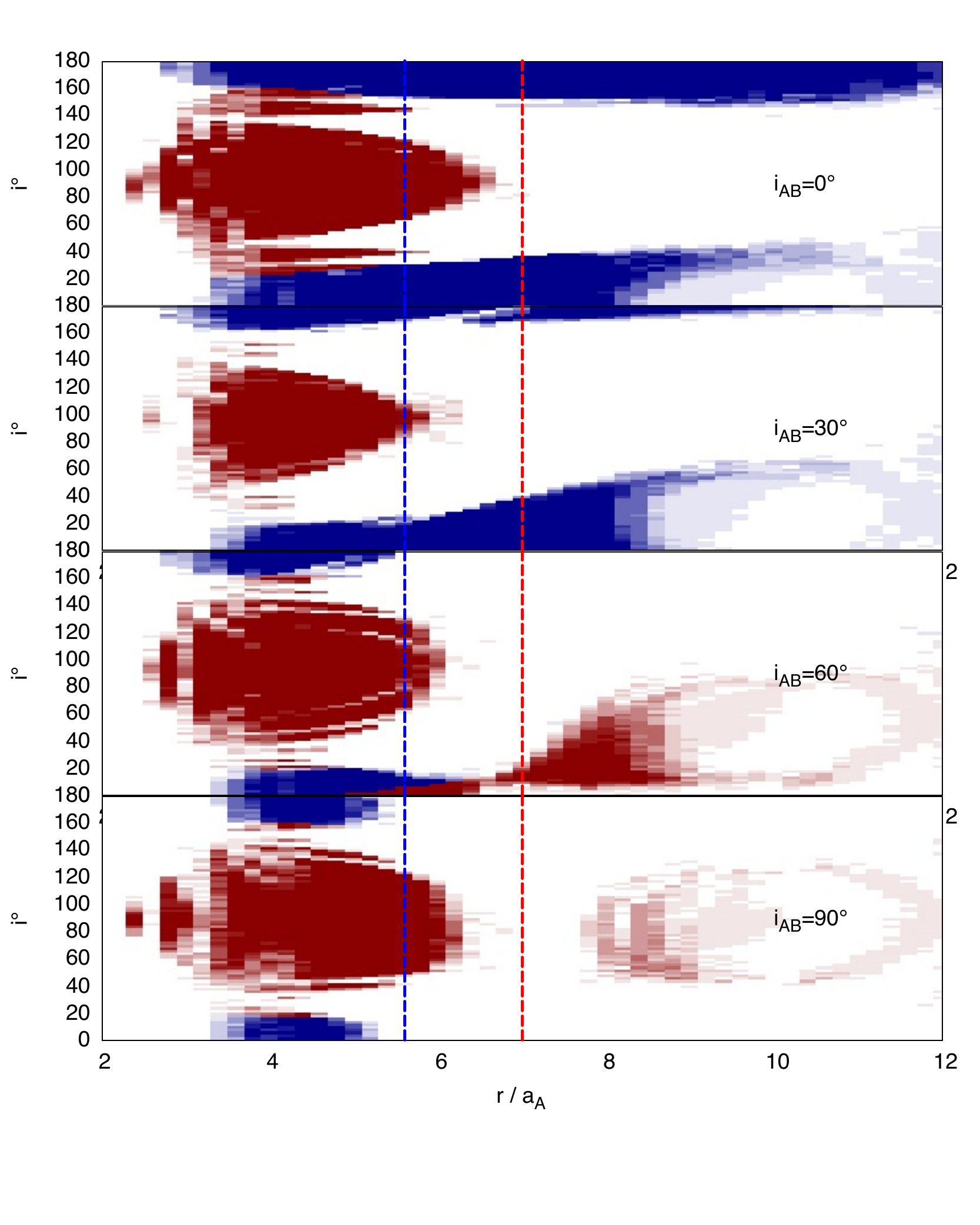}

\vspace{-20pt}
c)\,\,$e_{AB}=0.3$
\hspace{5cm}d)\,\,$f_A=0.1$ 

\includegraphics[width=8cm]{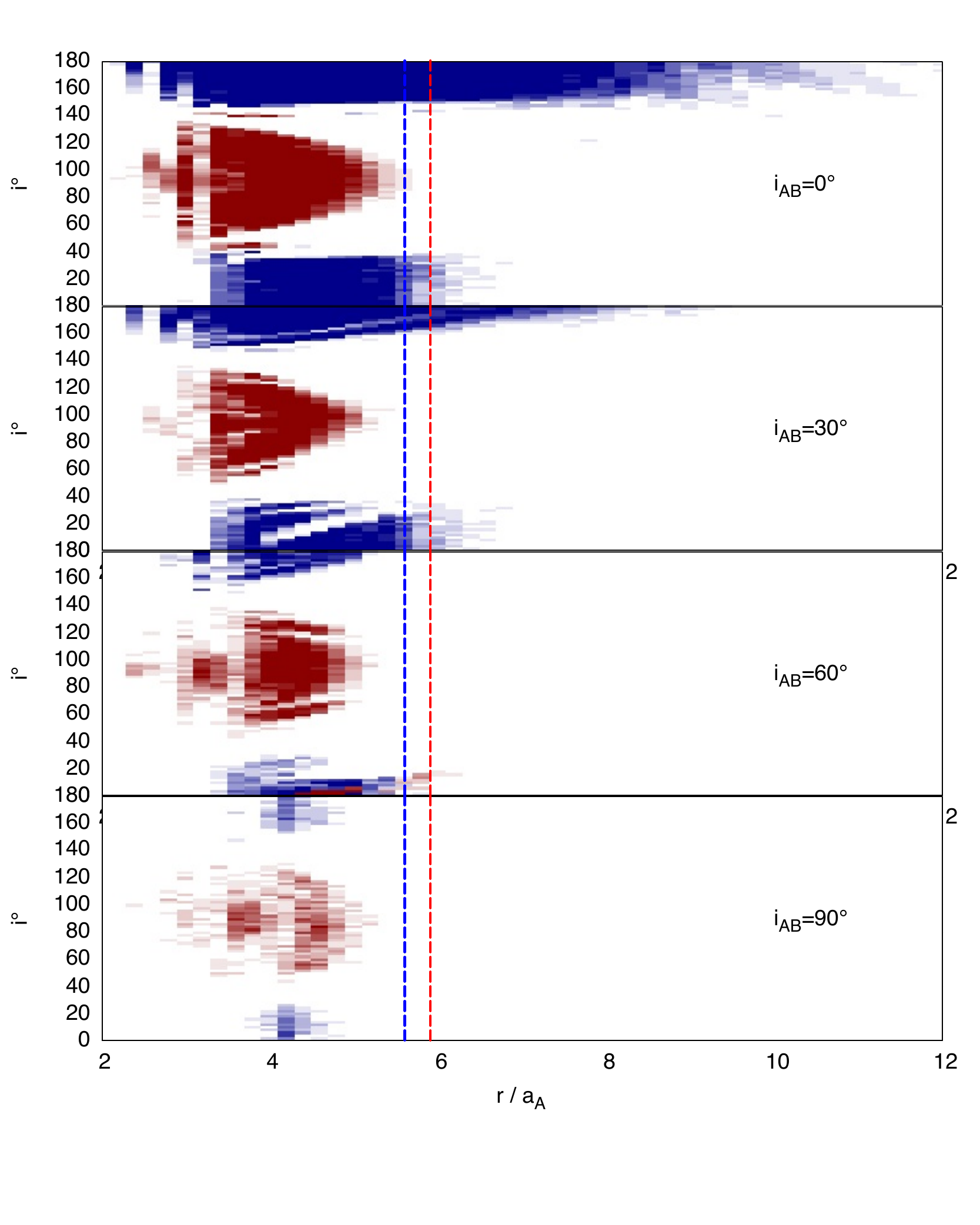}
\includegraphics[width=8cm]{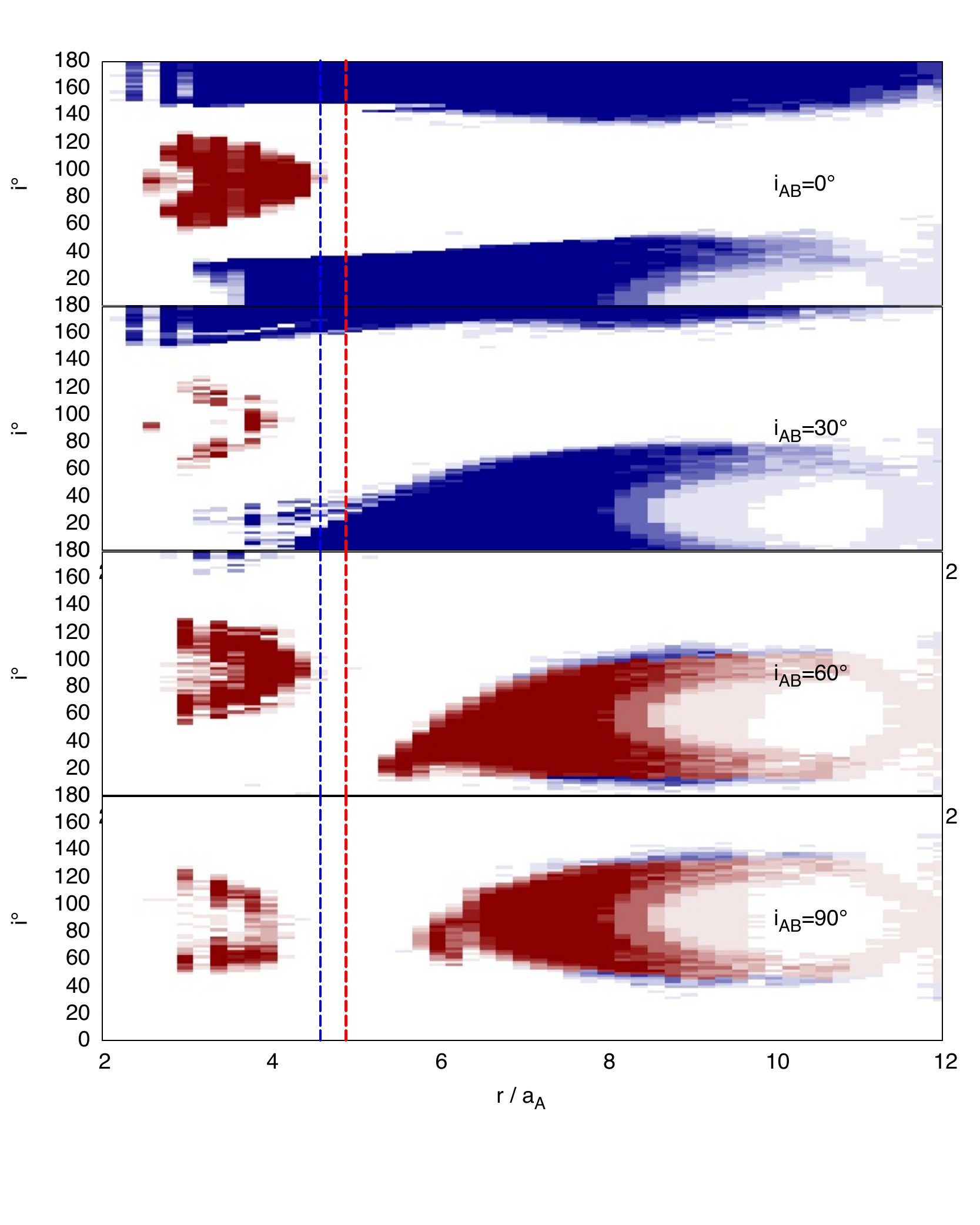}
\vspace{0.2cm}
\caption{Each panel is the same as the upper right panel in Fig.~\ref{stabilitymap1} with $\Omega_{AB}=90^\circ$ and the standard triple star parameters except one parameter is changed in each. The upper left panel has $e_A=0.2$, the upper right panel has $e_A=0.8$, the lower left panel has $e_{AB}=0.3$ and the lower right panel has $f_A=0.1$. }
\label{stab:p2} 
\end{centering} 
\end{figure*}

\subsection{Effect of the semi major axes ratio $a_{AB}/a_A$}

Figs.~\ref{stab:p1}a) and b) have the standard triple star parameters except the semi-major axes of the outer binary is $a_{AB}= 15$ and $60\,a_A$, respectively. 
For $a_{AB} =15\,a_A$ there are very few stable orbits. The value of  $R_{\rm L}$ from Equation~(\ref{eq:rc}) is about $4.0\,a_A$. Since this is very close to the inner unstable region of about $3\,a_A$, the librating region has nearly disappeared. The closer in companion has also completely destabilized the circulating region for higher mutual misalignments since the value for the circulating critical radius is  $R_{\rm C} =3.8\,a_A$.
On the other hand, for the larger semi major axes $a_{AB}=60\,a_A$, there is significantly more stability. The critical  $R_{\rm L}$ and $R_{\rm C}$ are about $9.1\,a_A$ and $8.7\,a_A$, respectively. The librating region is stable has greatly increased as are the region of circulating orbits at high mutual inclination.

\subsection{Effect of the companion mass}

Figs.~\ref{stab:p1}c) and d) have companion masses of $m_B=0.5$ and 2.0$\, m_A$, respectively.  Reducing the companion mass increases the stable librating region and in general makes for more stability.  For $m_B =0.5\, m_A$  the two critical radii are  $R_{\rm L}=6.9\,a_A$ and  $R_{\rm C} = 6.6\,a_A$ from Equation~(\ref{eq:rc}).  This is consistent with the librating regions shows in Fig.~\ref{stab:p1} c). Increasing the companion mass to $m_B=2\,m_A$ increases the influence from the companion and reduces the stable regions. Equation~(\ref{eq:rc}) gives  $R_{\rm L}=5.2\,a_A$ and  $R_{\rm C} = 4.8\,a_A$ for this case and one can see the reduced librating and circulating regions in the figure.

\subsection{Effect of the inner binary eccentricity}

Figs.~\ref{stab:p2}a) and b) have $e_A=0.2$ and $0.8$ respectively.  The smaller eccentricity has a slower libration time whereas the larger eccentricity has a faster libration time and so from Equation~(\ref{eq:rc}) for the $e_A=0.2$ case we have  $R_{\rm L}=4.8\,a_A$ and  $R_{\rm C}=5.6\,a_A$.  This is in good agreement with the figures  where the libration region has shrunk in comparison with our standard model but the circulating region stayed roughly the same.  For the $e_A=0.8$ case we get values of  $R_{\rm L}=7.0\,a_A$ and  $R_{\rm C}=4.6\,a_A$.  Again, this is in rough agreement with the figures  where the libration region has grown in comparison with our standard model but the circulating region stayed roughly the same.

\subsection{Effect of outer binary eccentricity}

Fig.~\ref{stab:p2}c) explores  the effect of the eccentricity of the outer binary.  From Equation~(\ref{eq:rc}) this gives values for $R_{\rm L}=5.9\,a_A$ and $R_{\rm C}=5.6\,a_A$, which are slightly reduced from our standard case.  The outer boundary has changed as the eccentricity of the companion allows for closer approach and stronger interactions.  Again this is consistent with stability of S-type orbits about a point mass replacing the inner binary \citep{Holman1999}, for which the unstable region begins at about $r = 6\,a_{A}$ for this configuration.

\subsection{Effect of the mass ratio of the inner binary}
\label{sec:ib}

Fig.~\ref{stab:p2}d) explores  the effect of the mass ratio of the inner binary.  Our standard case is equal mass but here we explore value of $f_A=0.1$.  From Equation~(\ref{eq:rc}) this gives values for $R_{\rm L}=4.9\,a_A$ and $R_{\rm C}=4.6\,a_A$ which are slightly reduced from our standard case. The libration region is reduced by even more then these numbers would suggest.  Also the inner boundary is farther out compared to our standard case, similar to that seen in \cite{Chen2020}.  These effects are probably due to higher order terms in the gravitational field coming from the unequal mass binary \citep{Li2014}.

\section{Discussion}
\label{sec:discuss}

The dynamics studied here have similarities with the dynamics of a test particle orbiting around an oblate planet that is orbiting around a star \citep[e.g.][]{Tremaine2009,lubow2025}. In that case, the test particle experiences competing quadrupole torques from an oblate planet and a distant stellar companion and the Laplace radius marks the transition between the dominance of the planetary quadrupole and the external tidal torque. The critical
radius $R_{\rm crit}$ that we derive in equation~(\ref{eq:rc})
plays an analogous role, separating regions where the
inner binary controls the test particle dynamics from those
where perturbations from the outer companion dominate.
However, in the present problem  the
eccentric inner binary supports both circulating and
librating particle orbits, including stable polar
configurations that precess about the binary eccentricity
vector \citep{Farago2010,Lubow2018}, whereas in the oblate planet case the orbits circulate around two fixed quadrupole fields an inner and an outer.  Furthermore, the quadrupole, 
in our case, is not fixed and evolves through interactions with the 
companion.  Despite these differences, the qualitative picture is 
similar with nearby test particles being stable and distant test 
particles subject to dynamics that can drive KL oscillations.

More generally, our work complements studies of
hierarchical secular dynamics in systems with multiple
perturbers. \citet{Hamers2015} and \citet{HamersLai2017}
showed that competing secular torques in hierarchical
quadruple systems can lead to complex and chaotic
evolution. While we consider the simpler restricted
problem of a massless circumbinary particle in a
hierarchical triple, the same physical principle
underlies the dynamics: long-term evolution is governed
by the competition between distinct quadrupolar
perturbations acting on different timescales.

Higher-order secular effects may also play a role in some
configurations
\citep[e.g.][]{Naoz2017}. For our standard model, the inner
binary is equal mass and the outer binary orbit is
circular, so the leading-order octupole terms vanish and
the quadrupole approximation is expected to dominate.
However, the unequal-mass models in Section~\ref{sec:ib} may be
influenced by higher-order effects, which could contribute
to the increased instability and reduced extent of the
librating region. A more detailed exploration of these
effects is left to future work.

Finally, our results complement studies of stability in
eccentric and mutually inclined planetary systems
\citep{Hadden2018,Bhaskar2024} by focusing specifically
on circumbinary orbits in hierarchical triples. The
existence of long-lived polar and highly inclined
configurations suggests that misaligned circumbinary
disks and planets may remain dynamically viable even in
the presence of strongly inclined stellar companions.

The results of this work have implications for the evolution of circumbinary gas disks with an outer stellar companion. The regions where particle orbits are stable describe the approximate extent of a stable gas disk, although hydrodynamical simulations are required to determine how pressure and viscosity can modify the stability boundaries.   The rings of the gas disk feel the same torque as the test particle although they communicate through viscosity and/or pressure \citep{Papaloizou1983,Papaloizou1995}. The disk can undergo nodal precession on a density weighted average timescale, similar to the particles. 

In addition, the effects of viscosity in a gas disk can lead to alignment towards a stationary state depending upon the type of nodal precession, circulating or librating. The disk aligns to coplanar or polar alignment relative to the inner binary \citep{Aly2015,Martin2017, Martin2018,Lubow2018,Zanazzi2018,Martin2019,Johnson2025}.  A finite-width circumbinary disk may span the critical radius, resulting in differential precession between its inner and outer regions. Such differential precession may lead to warping or disk breaking \citep[e.g.][]{Larwoodetal1996,Nixon2013,Nealon2016,Rabago2024}.  Our results show that while the radial extent of a polar gas disk may be small, it is not significantly affected by the inclination of the outer companion. Disks that are coplanar or retrograde coplanar may be much more radially extended if the outer companion is coplanar or retrograde coplanar, but they can be significantly truncated by a highly misaligned companion star.

Gas disks can also undergo global KL oscillations if they are sufficiently misaligned to the outer companion \citep{Martin2014,Fu2015,Lubow2017,Zanazzi2017}. This can lead to eccentricity growth within the disk, even if the disk is in a stationary orientation \citep{Martin2022}. This could have implications for planet formation in such disks such as where planets can form \citep{Aly2020,Martin2022kldust} and increase the possibility for disk fragmentation \citep{Fu2017}. We leave the exploration of such topics to future work.

\section{Conclusions}
\label{sec:conc}

We have investigated the stability of circumbinary test particle orbits in hierarchical  triple star systems where the inner binary is perturbed by a distant stellar companion. In this configuration the particle experiences competing torques from the inner binary and the outer companion. The inner binary drives nodal circulation around the binary angular momentum vector or nodal libration about the eccentricity vector.  The companion can induce additional precession and Kozai--Lidov (KL) oscillations that excite eccentricity and lead to instability. The long–term stability of circumbinary material is therefore governed by the competition between the inner–binary nodal precession time and the KL timescale imposed by the companion.

Using secular arguments we found an approximate critical radius $R_{\rm crit}$ at which the timescales for the nodal precession driven by the inner binary and KL oscillations driven by the outer binary are comparable. Inside this radius the inner binary dominates and circumbinary orbits may survive, while outside this region the companion drives KL cycles that typically destabilize the orbit, unless the orbit is close to coplanar or retrograde coplanar to the outer binary orbit. This simple timescale estimate predicts the outer boundary of the stable librating zone and agrees well with the $n$-body simulations.
 The size of this stable region increases for wider or less massive companions and for larger inner-binary eccentricity, and decreases for closer or more massive companions and for unequal inner-binary mass ratios.

The mutual misalignment between the inner and outer binaries plays a crucial role in the stability of circumbinary test particles, particularly for particle orbits that are coplanar (or retrograde coplanar) relative to the inner binary. Coplanar orbits can be highly unstable if the outer companion star is highly misaligned. On the other hand, the stability of orbits that are close to polar to the inner binary are not significantly affected by the mutual misalignment. Stable circumbinary orbits persist even in highly misaligned triples, including mutual inclinations up to $90^\circ$.

These results imply that circumbinary disks and planets can remain stable in triple systems over a substantial range of radii and inclinations, even when the outer companion is strongly misaligned. In particular, polar or highly inclined circumbinary configurations can be long–lived. The survival of such orbits provides a natural pathway for the formation of misaligned or polar circumbinary disks and planets, consistent with the growing number of observed misaligned gas disks and the possible evidence for polar circumbinary planets.

\section*{acknowledgements}

We thank the anonymous referee for many helpful comments
and suggestions that improved the clarity and scope of
this work.
We acknowledge support from the Nevada
Center for Astrophysics.
This research made use of the \texttt{REBOUND} N-body
package \citep{rebound} and the \texttt{WHFast}
integrator \citep{reboundwhfast}.
We acknowledge support from NASA through grant 80NSSC25K0346.


\bibliographystyle{aasjournal}
\bibliography{ct}

\end{document}